\documentclass[aps,prb,twocolumn,groupedaddress,amsmath,amssymb]{revtex4}
\usepackage{amssymb}
\usepackage{amsmath}
\usepackage{graphicx}
\usepackage{subfigure}
\usepackage{textcomp}
\usepackage{color}
\usepackage{amsfonts}
\usepackage{bbold}
\usepackage{dsfont}
\usepackage{epsfig}
\usepackage{hyperref}

\begin{document}

\title{Spin-pumping and inelastic electron tunneling spectroscopy in topological insulators}

\author{Aaron Hurley, Awadhesh Narayan, and Stefano Sanvito}
\affiliation{School of Physics and CRANN, Trinity College, Dublin 2, Ireland}

\date{\today}

\begin{abstract}
We demonstrate that a quantum spin Hall current, spontaneously generated at the edge of a two-dimensional topological
insulator, acts as a source of spin-pumping for a magnetic impurity with uniaxial anisotropy. One can then manipulate 
the impurity spin direction by means of an electrical current without using either magnetic electrodes or an external 
magnetic field. Furthermore we show that the unique properties of the quantum spin Hall topological state have profound effects 
on the inelastic electron tunneling spectrum of the impurity. For low current densities inelastic spin-flip events do not 
contribute to the conductance. As a consequence the conductance steps, normally appearing at voltages corresponding 
to the spin excitations, are completely suppressed. In contrast an intense current leads to spin pumping and generates a 
transverse component of the impurity spin. This breaks the topological phase yielding to the conductance steps.
\end{abstract}


\maketitle

{\it Introduction.---}The Quantum Spin Hall (QSH) state is a new topological phase of matter, which has attracted growing attention over the past 
few years. It is time-reversal invariant and characterized by both a bulk gap and gapless \emph{helical} edge states with opposite 
spins counter-propagating at a given edge. In a pathfinding work by Kane and Mele~\cite{Kane-Mele} it was shown that at low 
energy a QSH effect can be generated in a sheet of graphene subject to spin-orbit interaction. A topological invariant, $Z_2$, 
distinguishes such a QSH insulator from an ordinary one \cite{Kane-Mele2}. Unfortunately, spin-orbit interaction in graphene is 
too small to practically realize the QSH state at a realistic temperature, however a few other options are available. Bernevig {\it et al.} 
predicted that HgTe/CdTe quantum wells might exhibit this novel phase \cite{zhangHgTe0}, a prediction soon 
confirmed by experiments \cite{zhangHgTe1,zhangHgTe2}. Three-dimensional analogs of QSH state have also been found and 
are generically termed topological insulators (TIs). More recently, evidence for helical edge modes in InAs/GaSb quantum wells 
has also been found \cite{InAs}. Interestingly, both Silicene and two dimensional Ge have been predicted to exhibit QSH 
state at experimentally accessible temperatures \cite{yao-silicene,ezawa-silicene}, and there are theoretical proposals to realize 
the same in graphene by using non-magnetic adatoms \cite{franz-qsh}. These last two options would yield a material with a 
low-energy electronic structure identical to that proposed by Kane and Mele.

Given the peculiar spin structure of the QSH state it becomes natural to ask ourselves whether this can be used to manipulate 
magnetic objects~\cite{GPlatero}. In particular the question we address in this paper is whether spin-pumping at the single spin 
level can be achieved without using spin-polarized electrodes or an external magnetic field. In a nutshell we wish to propose an 
analog to the numerous recent investigations concerning spin-flip inelastic electron tunneling spectroscopy (SF-IETS) for magnetic 
adatoms on insulating surfaces \cite{Heinrich2004}, either in equilibrium or in spin-pumping conditions~\cite{Loth1}.

Here we demonstrate that a magnetic impurity deposited at the edge of a $Z_2$ TI and presenting a uniaxial magnetic anisotropy,
which makes it non-Kondo-active \cite{Maciejko}, can be manipulated by the QSH current. Furthermore we show that the topological
nature of the QSH state has profound consequences on the SF-IETS conductance spectrum. At low current intensity there is
a complete suppression of the conductance steps appearing at the critical biases characteristic of the activation of an inelastic 
spin-scattering channel~\cite{Heinrich2004}. In contrast, for currents large enough to produce spin-pumping the spin of the magnetic 
impurity is driven away from the anisotropy axis. This breaks the topological protection of the helical edge states and the conductance
steps reappear. Intriguingly, in this situation there is a strong dependence of the SF-IETS conductance spectrum on the bias polarity. 
Our calculations are conducted by using the non-equilibrium Green's function method for transport combined with a perturbative 
approach to spin-scattering from magnetic impurities \cite{Hurley1,Hurley2,Hurley3}.

\begin{figure}[t]
\centering
\resizebox{\columnwidth}{!}{\includegraphics[width=3cm,angle=-0]{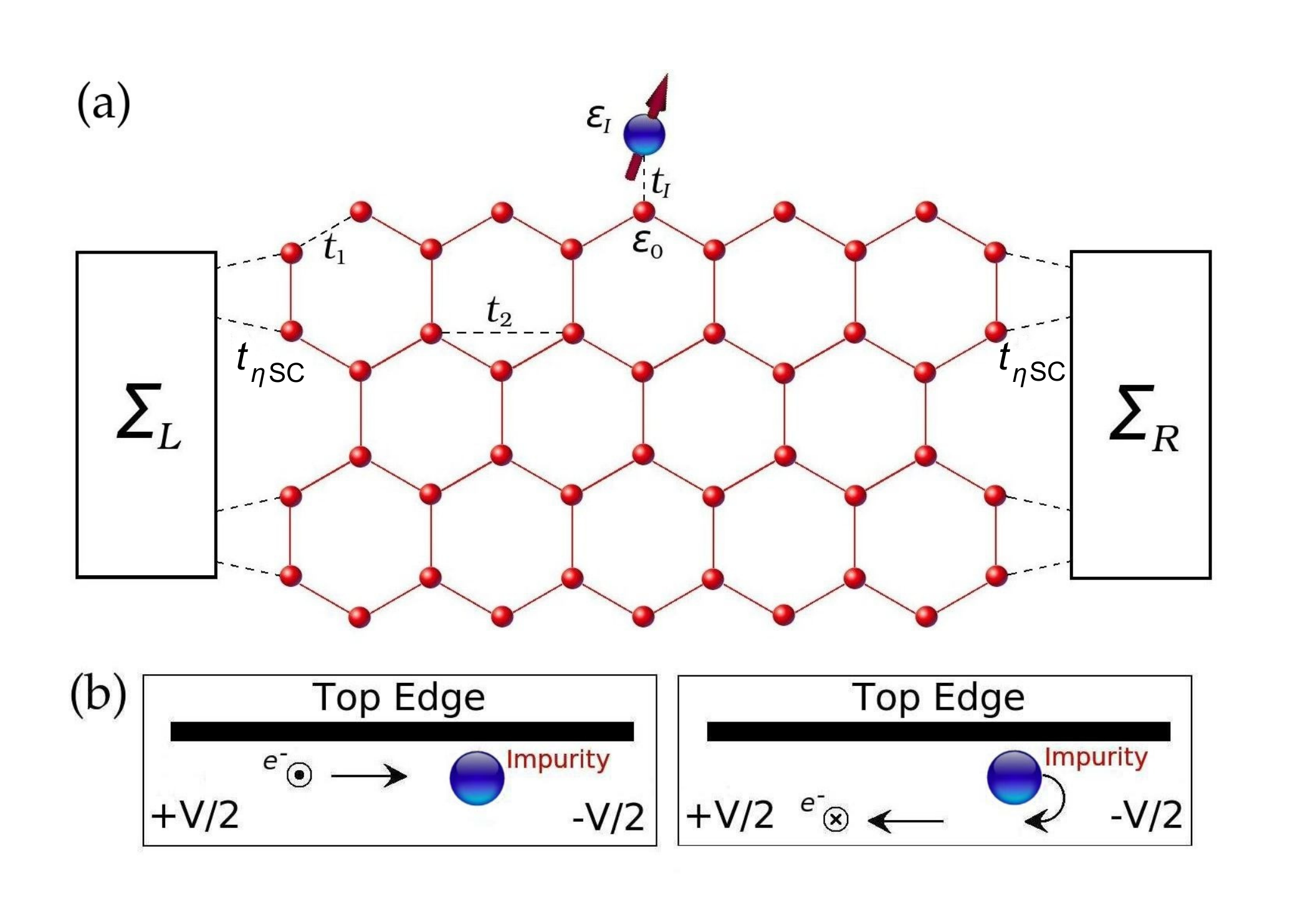}}
\caption{\footnotesize{Schematic representation of the device considered in this work (a), comprising a TI with honeycomb lattice 
structure and a magnetic impurity adsorbed at one of the two edges. In the transport calculations the electronic structures of the
electrodes is described by the self-energies, $\Sigma_{\eta}$ ($\eta=$~L, R). In panel (b) the cartoons describe a spin-flip scattering event.
A right-going electron with up spin direction (left panel) is inelastically back-scattered by the magnetic impurity. In the process both
the electron and the impurity spins are reversed (right panel). Note that, given the topological nature of the ribbon, spin-flip forbids 
electron transmission as the edge presenting a right-going spin-up state does not possess a right-going spin-down one.}}
\label{1}
\end{figure}

{\it Model and computational methods.---}The device that we have in mind is schematically presented in Fig.~\ref{1} and consists of two semi-infinite current/voltage 
electrodes sandwiching a $Z_2$ TI ribbon in which a magnetic impurity is absorbed at one of the two edges. Our working
hypothesis is that one can construct such a device with either a strong or a weak electronic coupling between the electrodes
and the ribbon, i.e. that the interface resistance can be engineered. The entire system is described at the tight-binding level 
and for the electrodes we use a simple square lattice with hopping parameter, $t_\eta$, ($\eta=$~L, R).

The TI ribbon has a honeycomb lattice structure with zig-zag edge geometry and it is described by the 2D Kane-Mele (KM) Hamiltonian, 
$H_{\mathrm{KM}}$, which reads
\begin{align}\label{hamiltonian-km} 
H_{\mathrm{KM}}=\varepsilon_0\sum_{i\alpha}\xi_i c^{\dagger}_{i\alpha}c_{i\alpha}&+t_1\sum_{\langle ij \rangle,\;\alpha}
c^{\dagger}_{i\alpha}c_{j\alpha}+\\ \nonumber
&+it_2\sum_{\langle\langle ij\rangle\rangle,\;\alpha\beta}\nu_{ij}c^{\dagger}_{i\alpha}[\sigma^z]_{\alpha\beta}c_{j\alpha}\:.
\end{align}
The first term describes a staggered sublattice potential with on-site energy $\varepsilon_0$ and $\xi_i$ being $\xi_i=+1$ for the A 
sublattice and $\xi_i=-1$ for the B one. Here $c^{\dagger}_{i\alpha}$ ($c_{i\alpha}$) creates (annihilates) an electron at site $i$ with 
spin $\alpha$. The second term is the nearest neighbour hopping with strength $t_1$ ($t_1$ sets the energy scale of the problem). 
Finally the third term, which drives the topological phase, is a second nearest neighbour hopping with strength $t_2$ 
($i=\sqrt{-1}$). This describes the coupling of the electrons orbital motion to their spins via the $z$-component of the Pauli 
matrices ($\sigma^z$). The parameter $\nu_{ij}$ is $+1$ for counter-clockwise hopping and $-1$ for clockwise. 

If we now attach an impurity at site $I$, the total electronic Hamiltonian will become
\begin{equation}
H_{\mathrm{el}}=H_{\mathrm{KM}}+\varepsilon_{I}\sum_{\alpha}c^{\dagger}_{I\alpha}c_{I\alpha}+t_I
\sum_{\langle Ii\rangle,\:\alpha}c^{\dagger}_{I\alpha}c_{i\alpha}\:,
\end{equation}
where in addition to $H_{\mathrm{KM}}$ one has the on-site potential of the impurity, $\varepsilon_I$, and the hopping between 
the impurity site $I$ and its neighbor $i$ on the honeycomb lattice (with strength $t_I$). Finally there are two other terms related to the
magnetic impurity spin, $\mathbf{S}$
\begin{equation}
H_{\mathrm{sp}}=DS^2_z\:;\;\;\;\;\;\;\;
H_{\mathrm{el-sp}}=J_{\mathrm{sd}}\sum_{\alpha\beta}c^{\dagger}_{I\alpha}[\boldsymbol{\sigma}]_{\alpha\beta}c_{I\beta}\cdot\mathbf{S}\:.
\end{equation}
The first, $H_{\mathrm{sp}}$, describes the uniaxial anisotropy (along $z$) with $D$ being the zero-field splitting parameter. The 
second, $H_{\mathrm{el-sp}}$, couples the current-carrying electrons to the local impurity spin with interaction strength 
$J_{\mathrm{sd}}$ ($\boldsymbol{\sigma}$ is a vector of Pauli matrices). This is usually known as the $s$-$d$ model for 
magnetism~\cite{sdmodel}.

Electron transport is investigated within the Landauer-B\"uttiker approach \cite{Buttiker} implemented by the non-equilibrium Green's function 
(NEGF) method~\cite{Datta}. The central quantity to evaluate is the retarded electronic Green's function for the scattering region (the TI ribbon) 
in presence of the electrodes, $G^r=[(E+i0^+)I-H_{\mathrm{el}}-\Sigma_\mathrm{L}-\Sigma_\mathrm{R}]^{-1}$, where $\Sigma_\eta$ 
($\eta=$~L, R) are the electrodes self-energy, which can be computed with standard techniques~\cite{Ivan}. These depend of the hopping
parameter between the ribbon and the electrodes, $t_{\eta\mathrm{SC}}$, whose magnitude sets the intensity of the current. 

When the conducting electrons couple to the impurity spin ($J_\mathrm{sd}\ne0$), the problem becomes intrinsically many-body in nature.
This is made treatable by constructing a perturbation theory in the electron-spin Hamiltonian, which allows us to incorporate the effect of the 
electron-spin interaction in an additional self-energy, $\Sigma_{\mathrm{int}}$. In this work we truncate the perturbation expansion to the 
second order~\cite{Hurley1,Hurley3} in both the electron and the impurity spin propagator. The latter contains information about the state 
of the magnetic impurity spins, through the population, $P_n$, of the eigenvectors of the spin-Hamiltonian, $H_\mathrm{sp}$. In particular 
it is possible to show that the $P_n$'s satisfy a master-equation of the form
\begin{equation}
\label{master}
\frac{dP_n}{dt}=\sum_{l}\Big[P_n(1-P_l)W_{ln}-P_l(1-P_n)W_{nl}\Big]+(P_n^0-P_n)/\beta\:,
\end{equation}
where the detailed expression for the transition rates, $W_{ln}$, can be found in Ref.~\cite{Hurley3}, and $\beta=k_\mathrm{B}T$ with 
$k_\mathrm{B}$ being the Boltzmann constant and $T$ the temperature. In Eq.~[\ref{master}] the populations $P_n^0$ are those of the
ground state. With this at hand we can compute the current, $I$, and hence by numerical derivative the conductance,
$G=\mathrm{d}I/\mathrm{d}V$.

{\it Results and discussion.---}We start our discussion by first looking at the transport properties of the ribbon in absence of the magnetic impurity. The relevant quantity here
is the spin-resolved total transmission coefficient along a particular edge \cite{Sheng}, which is given by
\begin{equation}
T^s_{\alpha\alpha'}(\varepsilon_\mathrm{F})=\mathrm{Tr}_{n_x}[\Gamma_\mathrm{L}G^r\Gamma_\mathrm{R}G^a]^{s}_{\alpha\alpha'}\:,
\end{equation}
where $\alpha$ is the spin index ($\alpha=\uparrow,\downarrow$), $s$ labels the edges ($s=$~top, bottom) and $G^a$ is the advanced
Green's function. The trace is over the number of atoms, $n_x$, along the given edge and the transmission coefficient is evaluated at the 
Fermi energy, $\varepsilon_\mathrm{F}$. As a matter of notation a $(n_x,n_y)$ ribbon contains $n_x$ atoms in the direction of transport 
and $n_y$ along the transverse one. When the Fermi level is fixed at the half-filling point the ribbon is insulating in the bulk, but presents 
edge topological protected states (here $t_{\eta}=t_1=1$, and $t_2=t_1/3$, which ensures that the KM Hamiltonian 
describes a QSH state). In this situation we find for a (11, 6) ribbon, $T^\mathrm{top}_{\uparrow\uparrow}=0.9$, 
$T^\mathrm{top}_{\downarrow\downarrow}=0.1$, $T^\mathrm{bottom}_{\uparrow\uparrow}=0.1$ and 
$T^\mathrm{bottom}_{\downarrow\downarrow}=0.9$. Such values demonstrate that the current along the QSH edges is spin-polarized,
although not completely because of the finite size of the ribbon. Calculations for a (7, 4) ribbon give us 
$T^\mathrm{top}_{\uparrow\uparrow}=0.85$, $T^\mathrm{top}_{\downarrow\downarrow}=0.15$, 
$T^\mathrm{bottom}_{\uparrow\uparrow}=0.15$ and $T^\mathrm{bottom}_{\downarrow\downarrow}=0.85$.

We now switch on the magnetic interaction between a $S=1$ local spin and a (11, 6) ribbon. In general we place the impurity 
at the centre of the edge and choose a coupling parameter, $t_I$, and an onsite energy, $\varepsilon_I$, to ensure that the density 
of states localized at the impurity site, $\rho_I(E)$, is approximately constant for energies, $E$, around the Fermi level (this ensures 
the convergence of the perturbation scheme \cite{Hurley1}). The exchange coupling, $J_{\mathrm{sd}}$, is chosen so that the 
perturbation parameter, $\rho_IJ_{\mathrm{sd}}$, is approximately 0.1. These conditions are satisfied for: 
$\varepsilon_I=J_{\mathrm{sd}}=t_1/2$ and $t_I=t_1/4$. The spin degeneracy is lifted by an axial anisotropy $D=-10^{-3}\:t_1$,
which corresponds to $D=-2.0$~meV, if the nearest neighbour hopping in the ribbon is fixed at a reasonable value of $t_1=2$~eV 
($k_\mathrm{B}T=0.05$). The uniaxial anisotropy gives us a degenerate ground state with the two spin states $|-1\rangle$ and 
$|+1\rangle$ separated from the first excited state $|0\rangle$ by $|D|$. As a result we do not expect a Kondo-like behaviour since 
no allowed transition between the degenerate ground state may occur. The second order perturbation expansion is then well justified. 
The values $t_{\eta}=4t_{\eta\mathrm{SC}}=t_1$ ensure that the spin system remains in equilibrium, i.e. in its ground state, 
throughout the spin-scattering process.

\begin{figure}[h]
\centering
{\includegraphics[width=8cm,angle=-0]{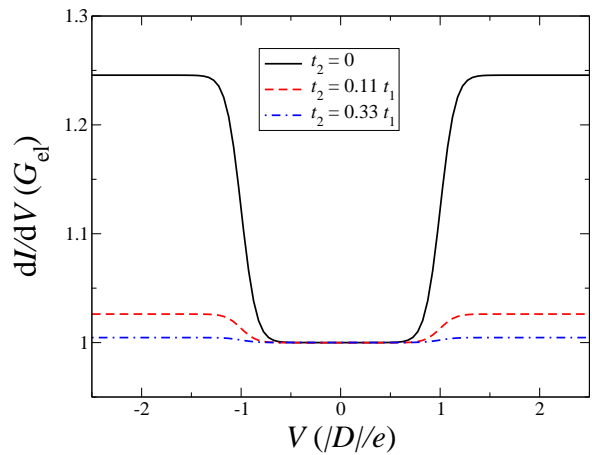}}
\caption{\footnotesize{(Color online) Sin-polarized IETS conductance spectrum for a TI (11, 6) ribbon with a $S=1$ magnetic impurity
attached at the upper edge. Note that the conductance step at the voltage characteristic of the inelastic excitation gets
suppressed as the $t_2$ parameter is increased, i.e. as the ribbon is brought well inside the topological region of the 
phase diagram.}}
\label{2}
\end{figure}

Figure~\ref{2} shows the calculated conductance spectra, $G(V)$, normalised to the $V=0$ conductance, $G_{\mathrm{el}}$, 
for three values of the parameter governing the QSH state, $t_2$. For $t_2=0$ there are no topologically protected edges and we 
observe the characteristic inelastic conductance step at a voltage $V=D/e$, when the transition from the ground state to $|0\rangle$ 
becomes possible ($e$ is the electron charge).  However, as $t_2$ is increased and we enter into the topological phase, we reveal 
a suppression of the inelastic contribution to the conductance, with an almost full suppression at the maximum value of $t_2=t_1/3$. 
The cartoon in Fig.~\ref{1}(b) helps to understand the mechanism for such a suppression. At a positive bias, the right-going current 
is up spin-polarized. This means that the $|-1\rangle\rightarrow|0\rangle$ transition scatters out spin-down electrons. These cannot 
propagate towards the right electrode since there is no right-moving spin-down state in the upper edge and, as a consequence, they 
are completely reflected. Hence, as spin-flip events can only lead to backscattered electrons, the inelastic channel does not contribute 
to the current. Note that the residual conductance increase in Fig.~\ref{2} for $t_2=t_1/3$ is simply due to the finite size of the ribbon.

We now investigate the possibility of manipulating the impurity spin direction. This is achieved by increasing the overall conductance, 
i.e. by increasing the average current density. When one works with an STM setup bringing the tip closer to the impurity \cite{Loth1}
does the job, while here our control parameter is the electronic coupling between the leads and the ribbon, $t_{\eta\mathrm{SC}}$.
As such all the calculations that follow have been performed with $t_{\eta\mathrm{SC}}=t_1$.

\begin{figure}[h]
\centering
\resizebox{\columnwidth}{!}{\includegraphics[width=5cm,angle=-0]{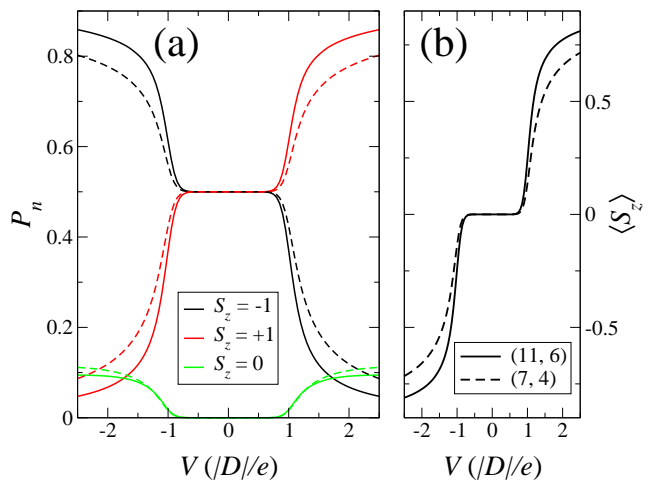}}
\caption{\footnotesize{(Color online) (a) Non-equilibrium population as a function of bias of the $S=1$ impurity spin states for a 
(7, 4) (broken lines) and a (11, 6) ribbon (solid lines). In panel (b) we show the average magnetization of the impurity for the same 
ribbons.}}
\label{3}
\end{figure}
The calculated populations of the various spin states are plotted as a function of bias in Fig.~\ref{3}(a) for both a (11, 6) and a (7, 4) ribbon.
A $S=1$ spin in a uniaxial anisotropy field and in thermal equilibrium with an electron bath presents an equal probability to occupy the 
$|+1\rangle$ and the $|-1\rangle$ states, i.e. for $V=0$ one has $P_{+1}=P_{-1}=1/2$. As soon as the bias is increased at and above
$|D|/e$, excitations to the $|0\rangle$ state become possible due to spin-flip back-scattering. In this case however the current is intense,
so that in between two scattering events the impurity spin does not have the time to relax back to the degenerate ground state. This means
that now a spin-up electron (the majority specie in the upper edge right-going channel) can also induce the transition 
$|0\rangle\rightarrow|+1\rangle$. The consequence is that the electronic current flowing at the upper edge, in virtue of its spin polarization 
and its intensity, produces a net flow of population between the two degenerate ground state, i.e.  for $V>+|D|/e$ one has $P_{+1}>P_{-1}$. 
In other words the impurity spin is driven by the current away from its uniaxial anisotropy axis. This can be fully appreciated by looking at 
Fig~\ref{3}(b), where we show the average magnetization $\langle S^z\rangle=\sum_mP_mS^z_{m}$ as a function of bias. Such spin-pumping 
is essentially identical to what happens for spin-polarized tips~\cite{Delgado} except that now one does not need either a magnetic electrode 
or an external magnetic field. Note that at a negative bias the effect is reversed, i.e. for $V<-|D|/e$ one has $P_{-1}>P_{+1}$, and that 
placing the impurity on the lower edge is equivalent to reversing the bias polarity.

\begin{figure}[t]
\centering
{\includegraphics[width=8cm,angle=-0]{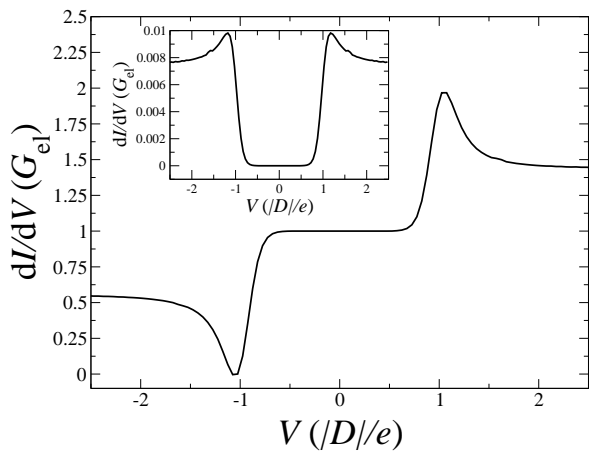}}
\caption{\footnotesize{Spin-polarized IETS conductance spectrum for a TI (11, 6) ribbon with a $S=1$ magnetic impurity
attached at the upper edge. In this case the current is intense and drives the impurity spin away from the uniaxial
anisotropy axis (see Fig.~\ref{3}). Notably now there is a step in the differential conductance at the voltage corresponding
to the inelastic transition $|\pm1\rangle\rightarrow|0\rangle$, The magnitude and sign of such step depends on the bias 
polarity. In the inset the inelastic contribution to the conductance.}}
\label{4}
\end{figure}

The effects of the spin-pumping on the shape of the conductance spectrum are finally presented in Fig.~\ref{4}. This time the $G(V)$ 
trace presents a step at the voltage corresponding to the $|\pm1\rangle\rightarrow|0\rangle$ transition, i.e. the electron transport
becomes sensitive to spin-flipping events. Such an appearance of the conductance step signals the suppression of the topological 
helical states induced by the transverse magnetization of the spin impurity \cite{AN}. Intriguingly, the magnitude and sign of the 
conductance step depends on the bias polarity. In particular we note that there is an inelastic contribution, which is symmetric with
respect to the sign of $V$, and always contributes to enhance the conductivity. In contrast the elastic contribution is anti-symmetric
with respect to the bias polarity, i.e. the elastic current increases for $V>|D|/e$ and decreases for $V<-|D|/e$. Placing the impurity 
on the opposite edge yields a mirror symmetric situation. This time the magnetization is driven toward $\langle S_z\rangle=-1$
($\langle S_z\rangle=+1$) and the conductance decreases (increases) for a positive (negative) bias voltage. Overall we can
summarize our results by noting that the sign of the change in the conductance trace at the onset voltage $|V|=|D|/e$ is proportional to 
$(\mathbf{v}\times\mathbf{\sigma})\cdot\langle\mathbf{S}\rangle$, where $\mathbf{v}$ is the group velocity of the edge state. Thus, the 
anti-symmetry of the conductance is related to the helicity of the edge state, $\mathbf{v}\times\mathbf{\sigma}$.

When one looks at the perturbative expansion of the conductance it can be realized that the term giving rise to the bias asymmetry is 
the magnetoresistive elastic term of the $s$-$d$ Hamiltonian. Its contribution to the self-energy reads 
\begin{equation}
[\Sigma_\mathrm{mag-elas}^{\gtrless}(E)]^{(2)} \propto -J_{sd}^{2}\sum_{mn}G_{0}^{\gtrless}(E\pm \Omega_{mn})\delta_{mn}P_{n}S_{mn}^{z},
\label{magelas}
\end{equation}
where $G_{0}^{\gtrless}$ is the electronic Green's function, $\Omega_{mn}$ the energy difference between the two spin states $|n\rangle$
and $|m\rangle$ and $S_{mn}=\langle m|S_z|n\rangle$ is the spin transition matrix elements. Since the terms includes $\delta_{mn}$ there is
only an elastic contribution ($\Omega_{nn}=0$), which involves no spin-flip events~\cite{Hurley3}. Such term is proportional to $S_{nn}^{z}$ 
and thus reverts its sign as the direction of impurity spin is reversed.
Note that the elastic and inelastic contributions to the conductance are calculated by partitioning the current into two parts, obtained respectively
from the elastic and inelastic energy-dependent self-energies. These, however, are evaluated from the same self-consistent electronic Green's 
function, meaning that the elastic and inelastic contributions are not completely disentangled. As such, it should not be surprising that the 
on-set of inelastic scattering is evident also in the elastic contribution to the conductance.

\begin{figure}[ht]
\centering
{\includegraphics[width=8cm,angle=-0]{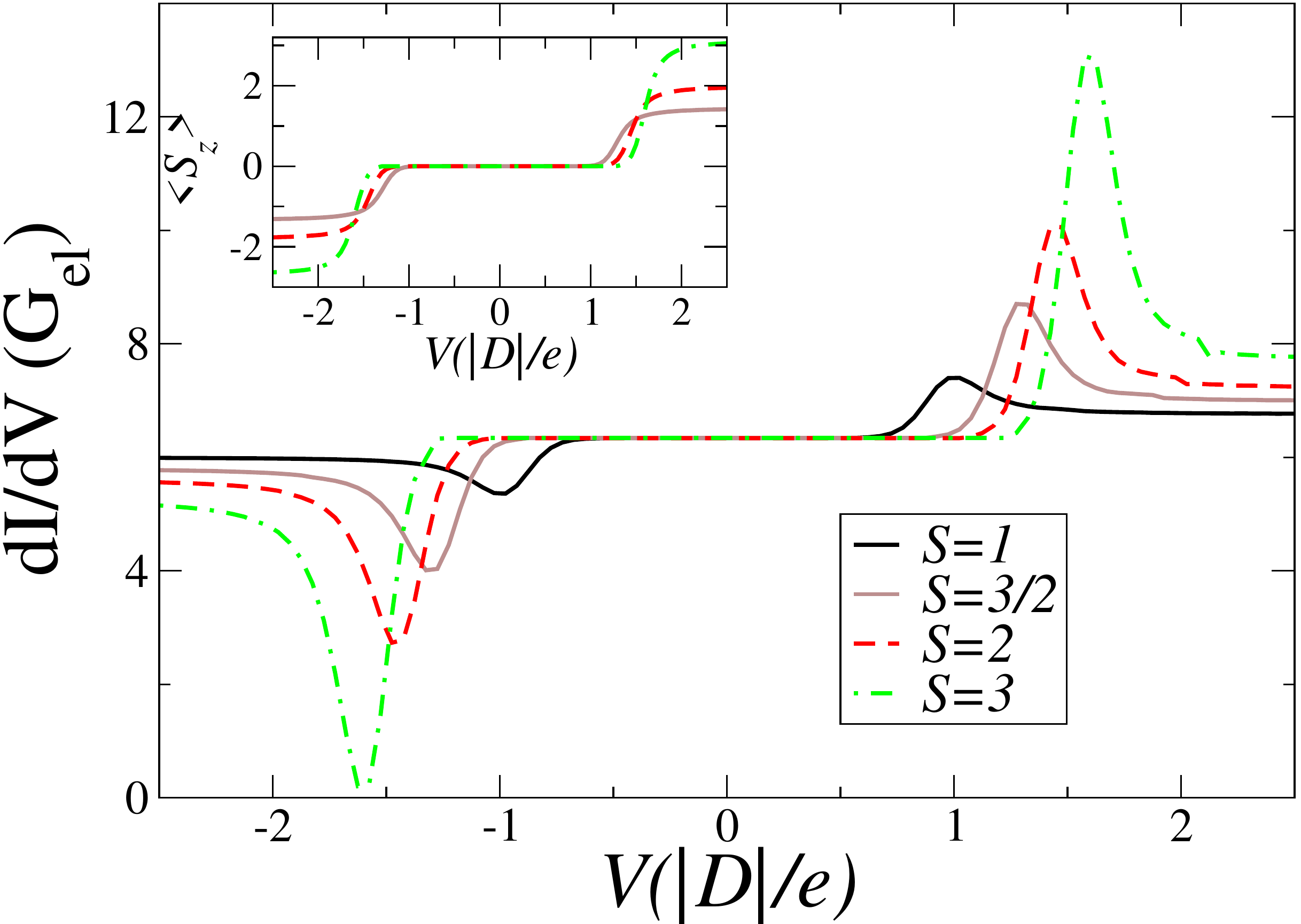}}
\caption{\footnotesize{(Color online) Spin-polarized IETS conductance spectrum for a TI (11, 6) ribbon incorporating a magnetic impurity with
various spin ($S=1, 3/2, 2, 3$) attached at the upper edge, in the intense current regime. The step in the differential conductance 
increases in magnitude with increasing the spin value of the adatom. Note that the spectra have been aligned vertically for clarity of 
comparison. The inset shows the average magnetization of the impurity for different values of $S$. Note that spin pumping persists for 
the larger values of the impurity spin.}}
\label{5}
\end{figure}

Finally we investigate how the conductance profile changes as we increase the value, $S$, of the total spin of the magnetic impurity. This
is done by rescaling the anisotropy parameter $D$ and the electron-spin coupling strength $J_\mathrm{sd}$ respectively to $D/|S|$ 
and $J_{\mathrm{sd}}/|S|$, so that the effective interaction strength and the total spin anisotropy do not change in the comparison. 
Our results are plotted in Fig.~\ref{5}. The figure reveals that, as the total spin increases, the height of the differential conductance steps 
gets larger. In the figure we also plot the average magnetization as a function of bias, which indicates that the spin pumping is present 
for larger values of $S$ and that for voltages exceeding the energy of the inelastic transition the average value of $S_z$ approaches 
its maximum value.

In conclusion we have demonstrated that a QSH current flowing at the edge of a $Z_2$ TI can be used to manipulate the
spin of a magnetic impurity. This does not require either an external magnetic field or magnetic electrodes, i.e. it allows one 
to implement spintronics without magnetism. Importantly the fingerprint of the manipulation can be found in the conductance
profile themselves, making SF-IETS a tool for preparing, manipulating and reading a quantum spin in the solid state.

{\it Acknowledgments.---}This work is supported by Science Foundation of Ireland (AH, grant No. 07/IN.1/I945) and Irish Research Council for 
Science, Engineering and Technology (AN). Computational resources have been provided by Trinity Center for High 
Performance Computing. We thank Cyrus Hirjibehedin, Magnus Paulsson and Nadjib Baadji for illuminating discussions.


\small
\begin{thebibliography}{}

\bibitem{Kane-Mele}
C.L.~Kane and E.J.~Mele, Phys. Rev. Lett \textbf{95}, 226801 (2005).

\bibitem{Kane-Mele2}
C.L.~Kane and E.J.~Mele, Phys. Rev. Lett \textbf{95}, 146802 (2005).

\bibitem{zhangHgTe0}
B.A.~Bernevig, T.L.~Hughes and S.C.~Zhang, Science {\bf 314}, 1757 (2006).

\bibitem{zhangHgTe1}
M.~K\"{o}nig, S.~Wiedmann, C.~Br\"{u}ne, A.~Roth, H.~Buhmann, L.W.~Molenkamp, X.L.~Qi and S.C.~Zhang, Science {\bf 318}, 766 (2007).

\bibitem{zhangHgTe2}
A.~Roth, C.~Br\"{u}ne, H.~Buhmann, L.W.~Molenkamp, J.~Maciejko, X.L.~Qi and S.C.~Zhang, Science {\bf 325}, 294 (2009).

\bibitem{InAs}
I.~Knez, R.R.~Du and G.~Sullivan, Phys. Rev. Lett \textbf{107}, 136603 (2011).

\bibitem{yao-silicene}
C.-C.~Liu, W.~Feng and Y.~Yao, Phys. Rev. Lett \textbf{107}, 076802 (2011).

\bibitem{ezawa-silicene}
M.~Ezawa, New Journal of Physics \textbf{14}, 033003 (2012).

\bibitem{franz-qsh}
C.~Weeks, J.~Hu, J.~Alicea, M.~Franz and R.~Wu, Phys. Rev. X {\bf 1}, 021001 (2011).

\bibitem{GPlatero}A.M.~Lunde and G.~Platero, Phys. Rev. B \textbf{86}, 035112 (2012).

\bibitem{Heinrich2004}
A.J.~Heinrich, J.A.~Gupta, C.P.~Lutz and D.M.~Eigler, Science {\bf 306}, 466 (2004).

\bibitem{Loth1}
S.~Loth, C.P.~Lutz and A.J.~Heinrich, Nature Physics {\bf 6}, 340 (2010).

\bibitem{Maciejko}
J.~Maciejko, Phys. Rev. B \textbf{85}, 245108 (2012).

\bibitem{Hurley1}
A.~Hurley, N.~Baadji and S.~Sanvito, Phys. Rev. B \textbf{84}, 035427 (2011).

\bibitem{Hurley2}
A.~Hurley, N.~Baadji and S.~Sanvito, Phys. Rev. B \textbf{84}, 115435 (2011).

\bibitem{Hurley3}
A.~Hurley, N.~Baadji and S.~Sanvito, Phys. Rev. B \textbf{86}, 125411 (2012).

\bibitem{sdmodel}K. Yosida, {\it Theory of Magnetism} (Springer-Verlag, Berlin, 1998)

\bibitem{Buttiker}
M.~Buttiker, Phys. Rev. B \textbf{38}, 9375 (1988).

\bibitem{Datta}S.~Datta, {\it Electronic Transport in Mesoscopic Systems} (Cambridge University Press, Cambridge, 1997)

\bibitem{Ivan}I.~Rungger and S.~Sanvito, Phys. Rev. B {\bf 78}, 035407, (2008)

\bibitem{Sheng}
L.~Sheng, D.N.~Sheng, C.S.~Ting, and F.D.M.~Haldane, Phys. Rev. Lett \textbf{95}, 136602 (2005).

\bibitem{Delgado}
F.~Delgado and J.~Fernandez-Rossier, Phys. Rev. B \textbf{82}, 134414 (2010).

\bibitem{AN}A.~Narayan and S.~Sanvito, Phys. Rev. B {\bf 86}, 041104(R) (2012).



\end{thebibliography}
\end{document}